\RequirePackage{setspace}
\PassOptionsToPackage{bookmarks=false}{hyperref}
\documentclass[article]{siamart1116} 

\usepackage{algorithmicx}
\usepackage{amsmath}
\usepackage{amssymb}
\usepackage{algorithm}
\usepackage{algpseudocode}
\usepackage[title]{appendix}
\usepackage[english]{babel}
\usepackage{booktabs}
\usepackage{graphicx}
\usepackage[utf8]{inputenc}
\usepackage{mathtools}
\usepackage{multirow}
\usepackage{siunitx}
\usepackage{tcolorbox}
\usepackage{varwidth}
\usepackage{verbatim}
\usepackage[margin=1.0in]{geometry}
\usepackage{tikz}
\usepackage{url}
\usepackage{wrapfig}

\usepackage{cite}

\usepackage{setspace} 
\usepackage{hyperref}
\usepackage{cleveref}

\numberwithin{theorem}{section}
\newcommand{\TheTitle}{Stochastic Block Models are a Discrete Surface Tension}
\newcommand{\TheAuthors}{Z. M. Boyd, M. A. Porter, and A. L. Bertozzi}
\newcommand{\TheShortTitle}{SBM is discrete surface tension}
\headers{\TheShortTitle}{\TheAuthors} 
\title{{\TheTitle}\thanks{Submitted to the editors DATE.}}
\author{\TheAuthors}
\author{
	Zachary M. Boyd\thanks{Department of Mathematics, UCLA, Los Angeles, CA (\email{zach.boyd@math.ucla.edu}).}
	\and
	Mason A. Porter\thanks{Department of Mathematics, UCLA, Los Angeles, CA (\email{mason@math.ucla.edu}).}
	\and
	Andrea L. Bertozzi\thanks{Department of Mathematics, UCLA, Los Angeles, CA (\email{bertozzi@math.ucla.edu}).}
}

\DeclareMathOperator{\const}{const}

\DeclareMathOperator*{\argmin}{argmin}
\DeclareMathOperator*{\argmax}{argmax}
\DeclareMathOperator{\Div}{div}

\DeclareMathOperator{\grad}{\nabla}
\DeclareMathOperator{\diag}{diag}

\newcommand{\nhat}{\hat{n}}
\newcommand{\Cut}{\mathrm{Cut}_{g,A}}
\newcommand{\w}{\omega}
\newcommand{\al}{\alpha}
\newcommand{\be}{\beta}
\newcommand{\vol}{\mathrm{vol}_{g,A}}
\newcommand{\sumab}{\sum_{\alpha,\beta}}

\newcommand{\dx}{\mathrm{d}x}
\newcommand{\dt}{dt}
\newcommand{\R}{\mathbb{R}}

\graphicspath{{./images/}}

\allowdisplaybreaks 
\everymath{\displaystyle}

\crefname{exp}{Expression}{Expressions}
\Crefname{exp}{Expression}{Expressions}
\crefname{app}{Appendix}{Appendices}
\Crefname{app}{Appendix}{Appendices}

\begin{document}

\maketitle

\begin{abstract}
Networks, which represent agents and interactions between them, arise in myriad applications throughout the sciences, engineering, and even the humanities. To understand large-scale structure in a network, a common task is to cluster a network's nodes into sets called ``communities", such that there are dense connections within communities but sparse connections between them. A popular and statistically principled method to perform such clustering is to use a family of generative models known as stochastic block models (SBMs). In this paper, we show that maximum likelihood estimation in an SBM is a network analog of a well-known continuum surface-tension problem that arises from an application in metallurgy. To illustrate the utility of this relationship, we implement network analogs of three surface-tension algorithms, with which we successfully recover planted community structure in synthetic networks and which yield fascinating insights on empirical networks that we construct from hyperspectral videos.

\end{abstract}







\section{Introduction}

The study of networks, in which nodes represent entities and edges encode interactions between entities \cite{newman_book}, can provide useful insights into a wide variety of complex systems in myriad fields, such as granular materials~\cite{papa2018}, disease spreading~\cite{rom-review2015}, criminology~\cite{crime_networks}, and more. In the study of such applications, the analysis of large data sets --- from diverse sources and applications --- continues to grow ever more important. 

The simplest type of network is a graph, and empirical networks often appear to exhibit a complicated mixture of regular and seemingly random features~\cite{newman_book}. Additionally, it is increasingly important to study networks with more complicated features, such as time-dependence \cite{holme2015}, multiplexity \cite{kivela2014}, annotations \cite{newman-clauset2016}, and connections that go beyond a pairwise paradigm \cite{otter2017}. One also has to worry about ``features'' such as missing information and false positives~\cite{jure}.
Nevertheless, it is convenient in the present paper to restrict our attention to undirected, unweighted graphs for simplicity.

To try to understand the large-scale structure of a network, it can be very insightful to coarse-grain it in various ways \cite{porter2009,fortunato_hric,peixoto,rombach2017,rossi2015}. The most popular type of clustering is the detection of assortative ``communities,'' in which dense sets of nodes are connected sparsely to other dense sets of nodes \cite{porter2009,fortunato_hric}. A statistically principled approach is to treat community detection as a statistical inference problem using a model such as a stochastic block model (SBM)\cite{peixoto}. The detection of communities has given fascinating insights into a variety of applications, including brain networks \cite{Betzel2016}, social networks \cite{fblong}, granular networks~\cite{granular_networks}, protein interaction networks~\cite{protein_networks}, political networks \cite{Porter2005}, and many others.

One of the most popular frameworks for detecting communities is to use an SBM, a generative model that can produce networks with community structure \cite{fortunato_hric,peixoto}.\footnote{Networks that are generated from an SBM can also have other types of block structures, depending on the choice of parameters; see~\cref{sec:sbms} for details.} 
One uses an SBM for community detection by fitting an observed graph to a statistical model to attempt to infer the most probable community assignment for each node. SBMs can incorporate a variety of features, including degree heterogeneity \cite{karrer_newman}, hierarchical structure \cite{peixoto2014}, and metadata \cite{newman-clauset2016}. The benefits of an SBM approach include statistical defensibility, theoretical tractability, asymptotic consistency under certain conditions, definable transitions between solvable and unsolvable regimes, and theoretically optimal algorithms~\cite{peixoto,moore}. As reviewed in \cite{fortunato_hric}, there are numerous other approaches for community detection, and statistical inference using SBMs is a method of choice among many people in the network-science community. A recent empirical study compared several types of SBMs and other community-detection approaches on a variety of examples \cite{ghasem2018}.

Recently, Newman showed that one can interpret modularity maximization \cite{newman-girvan2004,newman2006pre}, which is still among the most popular approaches for community detection, as a special case of an SBM \cite{newman_sbm}. In another paper~\cite{hui}, it was shown that one can also interpret modularity maximization in terms of graph cuts and total-variation (TV) minimization. The latter connection allows the application of methods from geometric partial differential equations (PDEs) and $\ell^1$ minimization to community detection. This relationship also raises the possibility of formulating SBM maximum-likelihood estimation (MLE) in terms of TV.\footnote{Another recent paper~\cite{hein} used total variation for maximizing modularity, although it was not phrased in those terms.} 
In this paper, we develop such a formulation, and we also incorporate substantial new ingredients to do so. The principal one is the notion of surface tension as a generalization of total variation. Additionally, we need to examine an energy landscape that requires a novel splitting--merging heuristic to navigate it, whereas previous graph-TV methods have been able to rely on gradient descent to discover satisfactory optima. Moreover, the dynamical systems that arise in the present work differ from those in~\cite{hui,bertozzi_flenner_2012}, in that our modified Allen--Cahn (AC) and Merriman--Bence--Osher (MBO) schemes involve diffusion with all-to-all coupling in addition to coupling that arises from a potential well, balance terms, or thresholding.

The main result of the present work is the establishment of an equivalence between SBMs and surface-tension models from the literature on PDEs that model crystal growth. Crystal growth is an important aspect of certain annealing processes in metallurgy~\cite{mullins_1956,surface_tension_survey}. It is a consolidation process, wherein the many crystals in a metal grow and absorb each other to reduce the surface-tension energy that is associated to the interfaces between them. The various processes involved have been modeled from many perspectives, including molecular dynamics~\cite{md}, front tracking~\cite{ft}, vertex models~\cite{vm}, and many others. (See~\cite{surface_tension_survey} for a much more extensive set of references.) It has been observed experimentally that the interface between any two grains evolves according to motion by mean curvature~\cite{cssmith}. Because mean-curvature flow is related to gradient descent (in the $L^2$ inner product) of the TV energy \cite{rof_1992}, this leads naturally to formulations in terms of level sets~\cite{osher_sethian}, phase fields~\cite{phase_field}, and threshold dynamics~\cite{MBO_1992}. Although the interfaces follow mean-curvature flow, each different interface can evolve at a different rate, as there are different surface-tension densities between each pair of crystals.  In realistic cases, surface tensions are both inhomogeneous and anisotropic, and they require careful adaptation of standard mean-curvature-flow approaches~\cite{esedoglu_otto,jacobs}, especially for dealing with the topological challenges that arise at crystal junctions, which routinely form and disappear. 

Recently, Jacobs showed how to apply techniques from models of crystal growth to graph-cut problems from semisupervised learning~\cite{jacobs}. (See~\cite{auction} for additional related work.) Several other recent papers, which do not directly involve surface tension, have used ideas from perimeter minimization and/or TV minimization for graph cuts and clustering in machine learning \cite{si_rev}. Three of those papers are concerned explicitly with ideas from network science~\cite{hui,boyd_modularity,hein}.

Each community in a network is analogous to a crystal, and the set of edges between nodes from a pair of communities is akin to the topological boundary between a pair of crystals. The surface-tension densities correspond to the differing affinities between each pair of communities. To demonstrate the relevance of this viewpoint, we develop and test discrete analogs of surface-tension numerical schemes on several real and synthetic networks, and we find that straightforward analogs of the continuum techniques successfully recover planted community structure in synthetic networks and reveal meaningful structure in the real networks. We also prove a theoretical result, in terms of $\Gamma$-convergence, that one can meaningfully approximate the SBM MLE problem by smoother energies. Finally, we introduce three algorithms, which are inspired by work on crystal growth, that we test on synthetic and real-world networks.

Our paper proceeds as follows. In \cref{back}, we present background information about stochastic block models, total variation, and surface tension. In \cref{equiv}, we state and prove our main result, which establishes an equivalence between discrete surface tension and maximum-likelihood estimation via an SBM. In \cref{gamma}, we discuss three numerical approaches for performing SBM MLE: mean-curvature flow, $\Gamma$-convergence, and threshold dynamics. We discuss our results on both synthetic and real-world networks in \cref{empirical}. In \cref{conc}, we conclude and discuss our results. We give additional technical details in appendices.


\section{Background}\label{back}


\subsection{Stochastic Block Models (SBMs)}
\label{sec:sbms}

The most basic type of SBM has $N$ nodes and an assignment $g:\{1,\ldots,N\}\to\{1,\ldots,\hat{n}\}$ that associates each node with one of $\hat{n}$ sets. It also has an associated $\hat{n}\times\hat{n}$ symmetric, nonnegative matrix $\w$ that encodes the affinities between pairs of communities. One generates an undirected, unweighted graph as follows: for each pair of nodes, $i$ and $j$, we place an edge between them with probability $\w_{\al\be}$, where $\al$ and $\be$ denote the community assignments of nodes $i$ and $j$, respectively. Similar models have been studied and rediscovered many times~\cite{peixoto,fortunato_hric,mle_1,mle_2,mle_3,harary_frank,condon2001}. In the present paper, we use the SBM from~\cite{newman_sbm}.

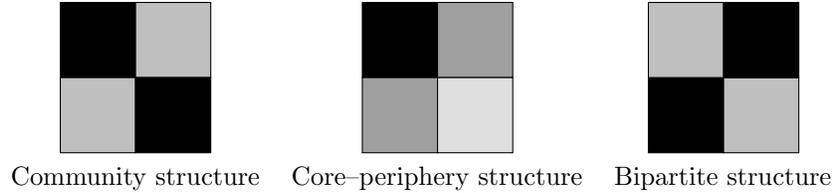
\begin{figure}
	\centering
	\begin{tabular}{ccc}
	\begin{tikzpicture}
		\path[draw=black,fill=black] (0,0) rectangle (1cm,-1cm);
		\path[draw=black,fill=black!50!white!50] (0,-1) rectangle (1cm,-2cm);
		\path[draw=black,fill=black!50!white!50] (1,0) rectangle (2cm,-1cm);
		\path[draw=black,fill=black] (1,-1) rectangle (2cm,-2cm);
	\end{tikzpicture}
	&
	\begin{tikzpicture}
		\path[draw=black,fill=black!100] (0,0) rectangle (1cm,-1cm);
		\path[draw=black,fill=black!75!white!50] (0,-1) rectangle (1cm,-2cm);
		\path[draw=black,fill=black!75!white!50] (1,0) rectangle (2cm,-1cm);
		\path[draw=black,fill=black!25!white!50] (1,-1) rectangle (2cm,-2cm);
	\end{tikzpicture}
	&
	\begin{tikzpicture}
		\path[draw=black,fill=black] (0,0) rectangle (1cm,1cm);
		\path[draw=black,fill=black!50!white!50] (0,1) rectangle (1cm,2cm);
		\path[draw=black,fill=black!50!white!50] (1,0) rectangle (2cm,1cm);
		\path[draw=black,fill=black] (1,1) rectangle (2cm,2cm);
	\end{tikzpicture}
	\\
	Community structure	&	Core--periphery structure	&	Bipartite structure
	\end{tabular}
	\caption{Examples of different connectivity patterns that one can generate using stochastic block models. Each panel corresponds to a different type of structure. In each panel, the upper-left and lower-right squares represent the density of connections between nodes in the same set, and the upper-right and lower-left squares represent the density of connections between nodes in different sets. Darker squares represent more densely connected sets of nodes. In (assortative) community structure, nodes are densely connected to other nodes in the same community but sparsely connected to nodes in other communities. In core--periphery structure, core nodes (as illustrated by the dark square in the upper left) are densely connected both to other core nodes and somewhat densely connected to peripheral nodes, but the latter predominantly have connections only to core nodes. In bipartite block structures, nodes in a set are more densely connected to nodes in other sets than to nodes in their own set. One can also model other structures, such as hierarchical and role-based structures, using SBMs. See~\cref{sec:sbms} for additional discussion. [This figure is inspired by a figure from~\cite{jeub2015}.]}
	\label{jeub_fig}
\end{figure}

There is considerable flexibility in the choice of $\w$, which leads in turn to flexibility in the SBMs themselves \cite{fortunato_hric,peixoto}. Three examples of $\w$, using $\hat{n}=2$, will help illustrate the diversity of possible block structures.
\begin{enumerate}
	\item If $\w_{11}=\w_{22} > \w_{12}$, one obtains traditional assortative community structure, in which nodes have a larger probability to be adjacent to nodes in the same community, instead of ones in different communities.
	\item If $\w_{11}=\w_{22} < \w_{12}$, nodes tend to associate more with nodes that are in other communities. As $\w_{12}\to 1,$ the graph becomes increasingly bipartite.
	\item If $\w_{11}>\w_{12} > \w_{22}$, there is a core--periphery (CP) structure: nodes from set $1$ are connected densely to many nodes, but nodes from set $2$ are connected sparsely to other nodes~\cite{csermely2013,rombach2017}.
\end{enumerate}
We illustrate these three examples in~\cref{jeub_fig}. To simplify our presentation, we refer to latent block structures as ``community structure,'' regardless of the form of the matrix $\w$.

The above SBM is not realistic enough for many applications, largely because each node has the same expected degree~\cite{karrer_newman}. To address this issue, one can suppose that one knows the degree sequence $\{k_i\}$ and then define connection probabilities to take this information into account. The easiest approach (see the discussion in~\cite{karrer_newman}) is to model the adjacency-matrix elements $A_{ij}$ (which is generated by the SBM) as Poisson-distributed with the parameter $\tilde{\omega}_{g_ig_j} := \w_{g_ig_j} \frac{k_ik_j}{2m}$, where $m$ is the number of edges in the associated network and $\w_{\al\be}$ is now allowed to take any value in $[0,\infty)$. This allows both multi-edges and self-edges. Although such edges can have important effects (including in configuration models)~\cite{Fosdick2016}, we neglect them for simplicity. Observe that the parameters $\w$, $k$, and $g$ are necessary and sufficient to specify $A$ as a random variable. In the present paper, we focus on the SBM that we described in this paragraph; it is known as a ``degree-corrected'' SBM~\cite{karrer_newman}.

Given an observed network, one can attempt to infer some sort of underlying community structure by statistical fitting methods. There are several ways to do this, including maximum-likelihood estimation (MLE), maximum a posteriori (MAP) estimation, and maximum marginal likelihood (MML) estimation. In MLE, one chooses the parameters $g$ and $\w$ under which an observed network is most probable (without using a prior), MAP estimation yields the most probable parameter configuration under a Bayesian prior, and MML estimation yields the best community assignment for each node individually by integrating out all of the other variables~\cite{moore,peixoto}. We use MLE, which is the simplest approach. In mathematical terms, the problem is to determine
\begin{equation}
	\argmax_{g,\w} P(A|g,\w)\,,
	\label{exp:mle}
\end{equation}
where $P$ is the probability density function. Because we determine the edges independently, $P$ is given by
\begin{equation*}
	P(A|g,\w) =\prod_{i\le j} P(A_{ij}|g,\w) =  \prod_{i\le j} P\left(A_{ij}\left|w_{g_ig_j}\frac{k_ik_j}{2m}\right.\right)\,.
\end{equation*}
We use a Poisson distribution, so
\begin{equation*}
	P(A_{ij}|\lambda) = 
	\begin{cases}
		\frac{\lambda^{A_{ij}}}{A_{ij}!}e^{-\lambda}	\,,&	i\ne j \,, \\
		\frac{\lambda^{A_{ij}/2}}{(A_{ij}/2)!}e^{-\lambda}\,,	&	i = j\,,
	\end{cases}
\end{equation*}
where the need for cases arises from our convention that $A_{ii} = 2$ if a self-edge is present. To solve~\cref{exp:mle}, one can equivalently maximize the logarithm of $P(A|g,\w)$. Conveniently, this changes the multiplicative structure into additive structure and allows us to drop irrelevant constants. The resulting 
problem is
\begin{equation}
  \argmax_{g,\w} \sum_{i,j}  \left[A_{ij} \log(\w_{g_ig_j}) - \w_{g_ig_j}\frac{k_ik_j}{2m}\right]\,.
	\label{exp:SBM_MLE}
\end{equation}
If $\w_{g_ig_j} = 0$, the quantity $A_{ij}\log(\w_{g_ig_j})$ is understood to be $0$ if $A_{ij}=0$ and $-\infty$ otherwise.

Common optimization heuristics for solving~\cref{exp:SBM_MLE} include greedy ascent~\cite{karrer_newman}, Kernighan--Lin (KL) node swapping~\cite{karrer_newman,kl_swapping}, and coordinate descent~\cite{newman_sbm}. As far as we are aware, the theory of these approaches has not received much attention.

In light of the extreme nonconvexity of the modularity objective function~\cite{good_2010} (which is known to be related to the planted-partition form of SBMs \cite{newman_sbm}), we expect that it is necessary to use multiple random initializations for any local algorithm. (Ideas from consensus clustering may also be helpful \cite{fortunato_hric}.)

Ways to elaborate the SBM of interest include incorporating overlapping and hierarchical communities \cite{peixoto2014,peixoto2015b}, generalizing to structures such as time-dependent and multilayer networks \cite{peixoto2015}, and incorporating metadata \cite{newman-clauset2016}. There are also Bayesian models and pseudo-likelihood-based methods~\cite{peixoto,amini}. We do not consider such embellishments in this paper, although we conjecture that it is possible to generalize our approach to some (and perhaps all) of these settings.


\subsection{Total Variation}

Consider a smooth function $f:\Omega\subset\R^d\to \R$ for some $d$. The total variation (TV) of $f$ is 
\begin{equation}\label{this} 
	|f|_{\text{TV}} = \int_{\Omega} |\grad f|\dx\,. 
\end{equation}	
For $d=1$, equation \eqref{this} describes the total amount of increase and decrease of the function $f$. If $f$ is smooth except for 
jump discontinuities along a smooth hypersurface $\Gamma$, one can interpret the derivative of $f$ in a generalized sense, yielding
\begin{equation*}
	|f|_{\text{TV}} = \int_{\R^d - \Gamma} |\grad f| \dx + \int_{\Gamma} \left|[f]\right| \dx\,,
\end{equation*}
where $[f]$ is the height of the jump across the discontinuity. The first integral uses a $d$-dimensional measure, and the second one uses a $(d-1)$-dimensional measure. In the particular case in which $d=2$ and $f$ is the characteristic function of some set $S$, we see that $|f|_{\text{TV}}$ is the perimeter of $S$. Similarly, when $d=3$, we obtain surface area.

Total variation is an important regularizer in machine learning. It is worth contrasting it with the Dirichlet energy $\int_\Omega |\grad f|^2\dx$, which has minimizers that satisfy $\Delta f = 0$, a condition that guarantees smoothness. However, minimizers of TV need not be smooth, as they can admit jump discontinuities. In image denoising, for instance, regularization using Dirichlet energy tends to blur edges to remove discontinuities, whereas a TV regularizer preserves the edges~\cite{rof_1992,candes_romberg_tao_2006}. 

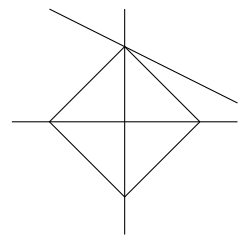
\begin{wrapfigure}{R}{.23\textwidth}
	\begin{tikzpicture}
		\draw (1,0) -- (0,-1) -- (-1,0) -- (0,1) -- (1,0);
		\draw (1.5,0.25) -- (0,1) -- (-1,1.5);
		\draw (1.5,0) -- (-1.5,0);
		\draw (0,1.5) -- (0,-1.5);
	\end{tikzpicture}
	\caption{Image of the $1$-norm unit ball and a line in the plane. The point on the line with the smallest $1$-norm is almost always on one of the axes.}
	\label{1-norm}
\end{wrapfigure}

Another use of TV energy is in relaxations, in which one can transform a nonconvex problem involving piecewise-constant constraints into a convex problem with the same minimizers~\cite{candes_romberg_tao_2006,egil_global}. 
A common heuristic explanation for this phenomenon (see~\cref{1-norm}) uses the shape of the $1$-norm unit ball. The simplest case is in two dimensions, where the $1$-norm ball is diamond-shaped, and minimizing the $1$-norm over certain domains (e.g.,~a line) gives a sparse solution, in the sense that most components of the solution vector are $0$. In this case, minimizing the $1$-norm, constrained to a line, is the same as minimizing the number of nonzero elements of the vector, subject to the same constraint. 

In the context of TV minimization, we take the $1$-norm of a function's gradient, rather than of the function itself. Therefore, instead of promoting sparsity of the function values, we promote sparse gradients, thereby incentivizing piecewise-constant minimizers for TV. Although our discussion is heuristic, note that the ideas therein can be treated rigorously~\cite{candes_romberg_tao_2006}.

Algorithmically, one can minimize TV using approaches such as phase-field models~\cite{phase_field} or threshold dynamics~\cite{MBO_1992}, both of which rely on the fact that the gradient descent (in the $L^2$ inner product) of TV is related to mean-curvature flow \cite{rof_1992}. The alternating-directions method of multipliers (ADMM)~\cite{admm} and graph-cut methods, such as the one in~\cite{boykov}, are also very effective at solving such problems. 

Thus far, we have restricted our discussion of TV to a continuum setting. There exist graph analogs of the mathematical objects --- gradients, measures, integrals, tangent spaces, divergences, and so on --- that one uses to define TV in a continuum setting.
For instance, for any function $f$ on the nodes of a graph and for any edge between nodes $i$ and $j$, the discrete derivative at $i$ in the direction $j$ is
\begin{equation*}
	\grad f(i,j) = f(j) - f(j)\,.
\end{equation*}
Using the inner products
\begin{align*}
	\langle f,g\rangle &= \sum_{i=1}^N f_ig_i \,, \\
	\langle \phi,\psi \rangle  &= \sum_{i,j}  A_{ij} \phi_{ij}\psi_{ij}
\end{align*}
on the spaces of functions on the nodes and edges, respectively, gives the divergence as the adjoint of the gradient:
\begin{equation*}
	(\Div \phi)_i = \sum_{j} A_{ij}\phi_{ji}\,.
\end{equation*}
In a continuum, an alternative definition of TV is 
\begin{equation} \label{variational_tv}
	|f|_{\text{TV}} = \sup \langle \Div \phi, f \rangle \,,
\end{equation}
where the supremum is over an appropriate set of test functions.
For a graph, \eqref{variational_tv} is equivalent to
\begin{equation*}
	|f|_{\text{TV}} = \frac{1}{2}\sum_{i,j}  A_{ij}|f(i) - f(j)| \,.
\end{equation*}
See~\cite{van_gennip_2014,gilboa_osher_2008} for a detailed justification of these definitions.

Some methods for graph clustering (e.g., see~\cite{luxborg_2007}) rely on the combinatorial graph Laplacian $L = \diag(k) - A$, which is a discrete analog of the continuum Laplacian $\Delta$. The continuum Laplacian arises in solutions to constrained optimization problems that involve the Dirichlet energy, so it is reasonable to expect minimizers of energies that involve the combinatorial graph Laplacian to have analogous properties to minimizers of the Dirichlet energy. Indeed, minimizers that arise from graph spectral methods are usually smooth\footnote{In this context, ``smooth'' entails that a value varies gradually along edges in a graph (i.e., adjacent nodes have similar values), although this is not a strict mathematical meaning. Minimizers of Dirichlet-type energies on graphs normally have this type of smoothness property, whereas minimizers of graph TV energies often have sharp interfaces between sets of nodes (which occurs, for example, if a solution must have a value of either $0$ or $1$).}, instead of having sharp interfaces, so one needs to threshold them in some way. Such thresholding is a major source of difficulties for attempts to obtain theoretical guarantees about the nature of minimizers after thresholding. By contrast, methods that use graph TV can directly accommodate piecewise-constant solutions~\cite{egil_global}, which do not require thresholding to give classification information. Several previous papers have exploited this property of TV on graphs~\cite{hui,bertozzi_flenner_2012,wei_zhu,bresson}.


\subsection{Surface Tension}

Very roughly, one can consider a metal object as being composed of a large number of crystals that range in size from microscopic to macroscopic~\cite{ashcroft_mermin}. Each crystal is a highly-ordered lattice; and there is a thin, disordered interface between crystals. The sizes and orientations of these crystals affect material properties, and one goal of annealing processes is to allow crystals to reorganize to produce a useful metal.

The potential energy of a crystal configuration is roughly 
\begin{wrapfigure}{L}{.3\textwidth}
	\includegraphics[width=.3\textwidth]{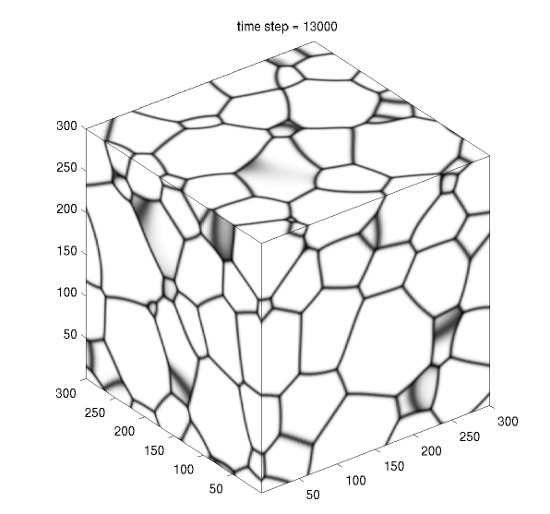}
	\caption{An example arrangement of crystals. The interfaces between pairs of crystals grow into each other according to motion by mean curvature. [This image is from Cenna/Wikimedia Commons/Public Domain~\cite{wiki}.]
	}
\end{wrapfigure}
\begin{equation}
	\sumab \sigma_{\al\be} \mathrm{Area}(\Gamma_{\al\be}) \,,
	\label{exp:cont_ste}
\end{equation}
where $\Gamma_{\al\be}$ is the interface between crystals $\al$ and $\be$, and $\sigma_{\al\be}$ is the surface-tension energy density between these crystals. Each $\sigma_{\al\be}$ is different, based on physical considerations that involve the exact offset between the orientations of the lattices in each pair of crystals. When prepared and heated appropriately, the individual crystals decrease~\cref{exp:cont_ste} by growing to consume their neighboring crystals. See~\cite{surface_tension_survey, jacobs, esedoglu_otto} for further background information.

In the study of SBMs, one can use TV to express~\cref{exp:SBM_MLE}, but we find a more natural formulation in terms of surface-tension energy (a related notion). Specifically, we exploit the appearance of surface area in~\cref{exp:cont_ste} to cast it as a TV problem. Mathematically, we model the metal as a region of space that is partitioned into $\hat{n}$ regions, corresponding to the crystals in the metal. Let $u^{\al}$ and $u^{\be}$, respectively, denote the characteristic functions of the regions $\alpha$ and $\beta$. Therefore,
\begin{equation*}
	\mathrm{Area}_{\al\be} = |u^{\al}|_{\text{TV}} + |u^{\be}|_{\text{TV}} - |u^{\al} + u^{\be}|_{\text{TV}}\,.
\end{equation*}
Each interface between two regions evolves according to mean-curvature flow. Consequently, the surface-tension flow is locally mean-curvature flow, except at the junction of three or more crystals~\cite{jacobs, esedoglu_otto}.
Because of this connection, one can use some of the ideas (such as phase-field and threshold-dynamics methods~\cite{esedoglu_otto})
from TV minimization to perform surface-tension minimization. When using threshold dynamics, it is possible to do theoretical analysis in the form of Lyapunov functionals, $\Gamma$-convergence, and descent conditions~\cite{jacobs}.


\section{An Equivalence Between SBM MLE and Discrete Surface Tension} \label{equiv}

We now present a mathematical result that connects SBM MLE and discrete surface tension.

\begin{proposition}
  Maximizing the likelihood of the parameters {$g \in \{1,\nhat\}^N$} (i.e., node assignments) and {$\w$} (i.e., affinities) in the degree-corrected SBM (see~\cref{exp:SBM_MLE}) is equivalent to minimizing
	\begin{equation}
		\sumab\left[W_{\alpha\beta} \Cut(\alpha,\beta)  + e^{-W_{\alpha \beta}} \frac{\vol(\alpha) \, \vol(\beta)}{2m}\right]\,, 
		\label{exp:cut_formulation}
	\end{equation}
	where $\Cut(\alpha,\beta) = \sum_{\substack{g_i = \alpha \\ g_j = \beta}} A_{ij}$, the volume term is $\vol(\alpha) = \sum_{g_i = \alpha} k_i$, and $W_{\al\be} = -\log\w_{\al\be}$ (so $W \in (-\infty,\infty]^{\nhat\times\nhat}$).
\end{proposition}

One immediately has the following well-posedness results. For a fixed $W$, the expression~\eqref{exp:cut_formulation} has a solution, because the state space over which one is minimizing is of finite cardinality. Furthermore, for fixed $g$, one can find the optimal $W=W(g)$ in closed form by differentiating with respect to each component of $W$ and setting the result to $0$~\cite{karrer_newman}. (We obtain a minimum of \ref{exp:cut_formulation} because it is concave up.) Therefore, the full problem, in which we allow $W$ to vary, also has a solution, because there are a finite number of candidate pairs $(g,W(g))$. Uniqueness is not guaranteed, because one can permute the community labels (and the corresponding entries in $W$) to obtain another minimizer. Another source of non-uniqueness is the possibility of symmetries in the underlying graph, which allows any optimizer to be converted to another optimizer by permuting the node labels. Continuous dependence of the minimizer on $A$ is automatic, because the set of possible values for $A$ is discrete.
 
The analogy with continuum surface tension is as follows. Graph cuts are analogous to surface area: given a domain in $\R^3$, one can superimpose a fine grid on space and count the number of edges that cross the boundary to estimate its surface area. In the limit of an 
infinitely fine grid, this estimate converges to the surface area under appropriate conditions~\cite{boykov_grid}. Similarly, graph volumes are analogous to continuum volumes.\footnote{For example, in a uniform square grid (ignoring boundaries), the sum of degrees for a set of nodes is proportional to the number of nodes, which in turn is roughly proportional to the area encompassed by filling in the squares associated with the selected grid nodes.}
The quantities $W_{\al\be}$ play the role of surface tensions $\sigma_{\al\be}$, so the first set of terms is analogous to~\cref{exp:cont_ste}. One can view the second set of terms as a soft volume constraint. A constraint is ``soft'' if violating it adds a finite penalty on an objective function, so minimizers usually approximately satisfy the constraint. Volume-constrained versions of~\cref{exp:cont_ste} have received a great deal of attention~\cite{surface_tension_survey,auction}.\footnote{As far as we are aware, our formulation of SBM MLE in terms of graph cuts and volumes is novel, although similar formulas have appeared previously in the literature (see, e.g.,~\cite{peixoto}).}

We now prove Proposition 3.1.

\begin{proof} (Proposition 3.1)

	In~\cite{newman_sbm}, it was shown that maximizing the log-likelihood of the parameters $g$ and $\w$ for a particular version of the degree-corrected SBM amounts to maximizing~\cref{exp:SBM_MLE}. Let $\Pi(G,\hat{n})$ be the set of partitions of the nodes of a graph $G$ (associated with an adjacency matrix $A$) into at most $\hat{n}$ sets. Substituting $W_{\alpha\beta} = -\log \w_{\alpha\beta}$ into~\cref{exp:SBM_MLE} gives
	\begin{equation*}	
	  \argmin_{\substack{W_{\alpha\beta} \in (-\infty,\infty] \\ g\in \Pi(G,\nhat)}} \sum_{i,j}  \left[A_{ij} W_{g_ig_j} + \frac{k_ik_j}{2m} e^{-W_{g_i g_j}}\right]\,. 
	\end{equation*}
	Rearranging the summations gives
	\begin{equation*}
	  \argmin_{\substack{W_{\alpha\beta} \in (-\infty,\infty] \\ g\in \Pi(G,\hat{n})}} \left[\sumab\sum_{\substack{g_i = \alpha \\ g_j=\beta}} A_{ij} W_{\alpha\beta} + \sumab  \sum_{\substack{g_i = \alpha \\ g_j = \beta}}\frac{k_ik_j}{2m} e^{-W_{\alpha \beta}}\right]\,, 
	\end{equation*}
	where the inner sums are over all nodes $i$ and $j$ such that $g_i=\alpha$ and $g_j=\beta$. Rearranging again gives
	\begin{equation*} 
	  \argmin_{\substack{W_{\alpha\beta} \in (-\infty,\infty] \\ g\in \Pi(G,\hat{n})}} \left[\sumab W_{\alpha\beta} \sum_{\substack{g_i = \alpha \\ g_j=\beta}} A_{ij}  + \sumab  e^{-W_{\alpha \beta}} \sum_{\substack{g_i = \alpha \\ g_j = \beta}}\frac{k_ik_j}{2m} \right]\,.
	\end{equation*}
	Using the definition of $\Cut$ in the first set of terms and summing over the $j$ index independently in the second set of terms gives
	\begin{equation*} 
	  \argmin_{\substack{W_{\alpha\beta} \in (-\infty,\infty] \\ g\in \Pi(G,\hat{n})}} \left[\sumab W_{\alpha\beta} \Cut(\alpha,\beta)  + \sumab  e^{-W_{\alpha \beta}} \sum_{g_i = \alpha}\frac{k_i}{2m}\vol(\beta)\right] \,.
	\end{equation*}	
	Finally, we sum over the $i$ index in the second set of terms to obtain
	\begin{equation}
	  \argmin_{\substack{W_{\alpha\beta} \in (-\infty,\infty] \\ g\in \Pi(G,\hat{n})}} \sumab \left[W_{\alpha\beta} \Cut(\alpha,\beta)  + e^{-W_{\alpha \beta}} \frac{\vol(\alpha)\,\vol(\beta)}{2m}\right]\,. 
	\end{equation}
\end{proof}

One difference between~\cref{exp:cut_formulation} and~\cref{exp:cont_ste} is that in~\cref{exp:cut_formulation}, one performs optimization over the $W_{\alpha\beta}$, whereas in~\cref{exp:cont_ste} (i.e., in a continuum), one ordinarily treats the surface-tension densities as fixed by the choice of material that one is modeling. Another difference is that the surface-tension coefficients in the graph setting can be any element of $(-\infty,\infty]$, subject only to the symmetry condition $W_{\alpha\beta} = W_{\beta\alpha}$ (see~\cref{back}). By contrast, for a continuum, one needs further restrictions to ensure well-posedness. Esedoglu and Otto~\cite{esedoglu_otto} proved the following sufficient conditions for well-posedness:
\begin{itemize}
	\item[(1)]{$\sigma_{\alpha\beta} \ge 0$  \,,}
	\item[(2)]{$\sigma_{\alpha,\alpha}  = 0$ \,,}
	\item[(3)]{$\sigma_{\alpha\gamma} + \sigma_{\gamma\beta} \ge \sigma_{\alpha\beta}$\,.}
\end{itemize}
In a graph setting, one can use a straightforward change of variables, $W_{\al\be} \to W_{\al\be}-\frac{1}{2}W_{\al\al} - W_{\be\be}$, to make $W$ satisfy requirement (2).\footnotemark 
\footnotetext{See~\cref{app:diag} for the change of variables, which causes the sum in~\cref{exp:cut_formulation} to instead be over all $\al\ne\be$, so that there are no ``internal'' surface tensions.}
In general, however, at least one of requirements (1) and (3) is not necessarily satisfied for a graph. Requirement (1) is false whenever some component of $W$ is negative; this occurs exactly when $\w$ has a component that is larger than $1$. In the continuum, requirement (3) has the interpretation of preventing ``wetting,'' where one phase can spontaneously appear between two others. In the graph case, such a restriction is unnecessary, because the number of points is fixed and finite, with no possibility of inserting points of another phase between two nodes.

The analogy of~\cref{exp:cut_formulation} with continuum surface tension is simplest for the case of assortative communities, although it is also relevant for other types of block structure. For disassortative blocks, rather than an energy cost from surface area, particles in one phase can achieve a lower energy by interacting with particles in a different phase. This leads to solutions that maximize surface area, and the evolutions that we will consider in~\cref{gamma} then involve backward diffusion. In the continuum case, this is ill-posed; however, a graph does not include arbitrarily small length scales, so ill-posedness does not cause a problem. Backward diffusion on graphs also appeared recently in~\cite{gilboa_diffusion} in the context of image processing, and it would be interesting to see if their techniques would be insightful in our context. For core--periphery structure, one phase has an energy penalty from interacting with itself but lowers its energy by interacting with the other phase. The other phase, however, prefers to interact with itself. As far as we are aware, such structures have not been studied previously in the literature on surface-tension models of crystal growth. For more complicated block structures, it is concomitantly more complicated to interpret the analogy with continuum surface tension. In a sense, one should view the SBM in~\cref{exp:cut_formulation} as a generalization of discrete surface tension, rather than as an analog. In the present paper, we emphasize applications to assortative community structure.


\section{Mean-Curvature Flow (MCF), \texorpdfstring{$\Gamma$}{Gamma}-Convergence, and Threshold Dynamics}
\label{gamma}

We now outline three algorithmic approaches that illustrate how one can use tools from surface-tension theory to solve SBM MLE problems. Our three algorithms are graph versions of mean-curvature flow (MCF), Allen--Cahn (AC) evolution, and Merriman--Bence--Osher (MBO) dynamics. In~\cref{empirical}, we will conduct several numerical experiments to demonstrate that these algorithms can effectively solve~\cref{exp:SBM_MLE}. We expect the performance of these algorithms to be good relative to other algorithms for SBM MLE, although a full evaluation of this claim is beyond the scope of our paper.
We have posted our code at \url{github.com/zboyd2/SBM-surface-tension}. In the next three subsections, we describe how we infer $g$ when $\w$ is fixed and finite. We then describe how to jointly infer $\w$ and $g$.


\subsection{Mean-Curvature Flow}

Surface-tension dynamics are governed by mean-curvature flow except at junctions. Intuitively, each point on a surface moves in the direction normal to the surface at a speed given by the mean curvature at that point. In the two-phase case, such dynamics have been well-studied, and there exist notions of viscosity solutions and regularity theory~\cite{mcf_notes}. In the multiphase case, the situation is much more complicated, especially because of the topological changes that can occur and the issue of defining the behavior at the junction of three or more phases. In two-phase surface-tension dynamics, it was shown in~\cite{boykov} that one can approximate the flow by solving a discrete-time minimizing-movements problem. Let $C_n$ be one of the two regions at time $n\, \dt$, where $\dt$ is the time step. To update $C_n$, one uses
\begin{equation}
	C_{n+1} = \argmin_C \left[\mathrm{Surface Area}(C) + \frac{1}{\dt} \int_{C_n \Delta C} \hat{\rho}(p,C_n) \mathrm{d}p\right],
	\label{eqn:cont_mcf}
\end{equation}
where
\begin{equation*}
	\hat{\rho}(p,C_n) = \inf_{x\in \partial C_n} \|x - p\|\,,
\end{equation*}
the operation $\Delta$ denotes the symmetric difference, and $\partial$ is the topological boundary operator. The idea behind this approach is, at each time step, to shorten the curve as much as possible without straying too far from the curve location at the previous time step. 

In the setting of graphs, a similar approach was developed in~\cite{van_gennip_2014}, where the mean-curvature flow was given by
\begin{equation}
	C_{n+1} = \argmin \left[\Cut(C,C^c) + \frac{1}{\dt}\sum_{i\in C_n\Delta C} \rho(i,\partial(C_n))\right] \,,
	\label{eqn:grf_mcf}
\end{equation}
the operation $\Delta$ is again the symmetric difference, and $\rho(i,\partial(C_n))$ is the shortest-path distance from node $i$ to the boundary of $C_n$. In this context, the boundary of a set of nodes is the set of nodes in $C_n$ with at least one neighbor in $C_n^c$ along with the nodes in $C_n^c$ that have at least one neighbor in $C_n$. We use the term \emph{boundary node} for any node that lies on the boundary. In the limit of small $\dt$,~\cref{eqn:grf_mcf} may still evolve, as opposed to the MBO scheme (which we use later), which becomes ``stuck'' when the time step is too small. Such evolution can still occur, because the penalty (associated with moving any node in $\partial(C_n)$) induced by the second set of terms in~\cref{eqn:grf_mcf} is $0$, regardless of the value of $\dt$. Conveniently, this implies for sufficiently small $\dt$ that the only acceptable moves at each time step are ones that are allowed to change only the boundary nodes themselves. This makes it possible to drastically reduce the search space when solving~\cref{eqn:grf_mcf}.

Because careful studies in the spirit of~\cite{van_gennip_2014} are not yet available for multi-way graph partitioning, we resort to a heuristic approach based on what is known for bipartitioning. Specifically, we are motivated by the situation in which time steps are sufficiently small that only boundary nodes can change their community assignment. Ideally, we wish to compute an optimal reassignment of all boundary nodes jointly to minimize~\cref{exp:cut_formulation}. To save computation time and facilitate implementation, we instead decouple the computations in the following manner. During a single time step, for each boundary node, we compute an optimal assignment of that node, assuming that all other nodes keep their assignment from the beginning of the time step. After this (but before the end of the time step), we assign each boundary node to its community, as computed previously in the time step. Because most nodes are boundary nodes\footnote{Recall that a node is a boundary node if it shares an edge with a node that lies outside of its own community, so most reasonable partitions of many real graphs have many boundary nodes. Additionally, because we initialize $g$ with nodes assigned to communities uniformly at random, most nodes are initially boundary nodes for most graphs.}
in our SBM-generated graphs, we find it both more efficient and easier to consider reassigning all nodes in each time step, rather than maintaining and referencing a separate data structure to track the boundary. 
For other networks and initialization techniques, such as in networks that arise from nearest-neighbor graphs with initialization from spectral clustering, it may be more efficient to loop over only the boundary nodes. (This idea aligns particularly with the spirit of mean-curvature flow.\footnote{
There are similarities between gradient-descent methods and greedy approaches, because both attempt to make locally optimal moves. Our decisions to move nodes such that each move is conditionally independent and to not track boundary nodes is also reminiscent of greedy approaches. Ultimately, which nodes are considered at each time step is an implementation detail, because only boundary nodes change assignment (at least in the assortative case that we emphasize in this paper). For larger graphs with more communities and fewer boundary nodes, it may be possible to increase efficiency by considering moves of boundary nodes only to neighboring communities rather than our present approach of considering moves of nodes to any community. At the scale of our examples (up to millions of edges), this implementation choice is not necessary.}) In \cref{alg:mcf}, we give pseudocode for this graph MCF procedure.

\begin{algorithm}
	\begin{algorithmic}
		\State Input $A$, $W$, $\hat{n}$\,.
		\State Initialize $g$ so that nodes are assigned to communities uniformly at random.
		\State Let $\mathrm{eW} = e^{-W}$\, be the entry-wise exponential of $W$\,.
		\While{not converged}

		\State Let $U_{i\alpha} = \delta_{g_i\al}$ for each $i,\alpha$, where $\delta$ is the Kronecker delta.
		\State Let $X = AU$\,. \quad  // Counts the number of neighbors that each node has in each community
		\State Let $\vol = (k^TU)$\,.
		\For{$a'=1$ to $\hat{n}$}

		\State Let $I_{a'}$ be the set of nodes that are currently assigned to community.
		$a'$\,.

		\For{a=1 to $\hat{n}$}

		\State Let $I$ be the indices $1, \ldots, \hat{n}$ aside from $a$ and $a'$\,.
		\State Let $\mathrm{Delta}(I_{a'},a)$ be given by the following formula:
		\begin{align*}
			\mathrm{Delta}(I_{a'},a)& = 2X(I_{a'},I)W(I,a) \\
			& - 2X(I_{a'},I)W(I,a') \\
			& + 2X(I_{a'},a)W(a,a) \\
			& - 2X(I_{a'},a)W(a,a') + 2X(I_{a'}, a')W(a,a') \\
			& - 2X(I_{a'},a')W(a',a') \\
			& + \frac{1}{2m}\left(  2k(I_{a'})\vol(\mathrm{eW}(:,a) - \mathrm{eW}(:,a'))\right. \\
			& + \left.k(I_{a'})^2( \mathrm{eW}(a,a) + \mathrm{eW}(a',a') - 2\mathrm{eW}(a,a') )  \right).
		\end{align*}

		\EndFor
		\EndFor

		\For{$i=1$ to $N$}
		\State $g_i = \argmin(\mathrm{Delta(i,:)})$\,. \quad // [Choose uniformly at random in case of a tie.]
		\EndFor
		\EndWhile
		\State Output $g$\,.
	\end{algorithmic}
	\caption{Modified graph mean-curvature flow (MCF) for SBM MLE~\cref{exp:SBM_MLE}.
	}
	\label{alg:mcf}
\end{algorithm}


\subsection{Allen--Cahn (AC) Evolution}

Another approach for studying MCF is approximation by a Ginzburg--Landau (GL) functional. This approach is popular due to its simple implementation and the existence of unconditionally-stable numerical methods~\cite{bertozzi_flenner_2012}. 

In the two-phase case, the GL functional is
\begin{equation}\label{gd}
	\int_\Omega  \left[\epsilon\left|\grad u\right|^2 + \frac{1}{2\epsilon} u^2(1-u)^2\right]\dx\,,
\end{equation}
where $u:\Omega\subset\R^{N}\to\R$ is a smooth function and $\epsilon$ is a small parameter.

 The $L^2$ gradient descent of the GL functional is
\begin{equation*}
	u_t = \epsilon \Delta u - \frac{1}{2\epsilon} \frac{\mathrm{d}}{\mathrm{d}u}\left[u^2(1-u)^2\right]\,,
\end{equation*}
which is the Allen--Cahn (AC) equation. The minimizers of the GL energy are predominantly (piecewise) constant, with $O(\epsilon)$-width transition layers between the constant regions. One can show that the GL energy $\Gamma$-converges to the TV energy as $\epsilon\to 0$, assuming that $\int_\Omega u\, \dx=\const$~\cite{modica}. Consequently, if $u_\epsilon$ is a minimizer of the constrained GL energy with parameter $\epsilon$ and the minimizers converge in $L^1$ as $\epsilon\to0$, then the accumulation point is a minimizer of the TV energy. 

In the setting of graphs, the first use of AC schemes for TV minimization was in~\cite{bertozzi_flenner_2012}. One can invoke the combinatorial graph Laplacian $L = \mathrm{diag}(k) - A$ to obtain a \emph{graph GL functional}
\begin{equation}
	U^TLU + \frac{1}{\epsilon} U^2(1-U^2)\,,
	\label{exp:f1}
\end{equation}
where $U$ is a function on the graph nodes (so it is an $N$-element vector) and $\epsilon$ is again a positive number. Expression~\eqref{exp:f1} $\Gamma$-converges to graph TV~\cite{van_gennip_2012}. 

In the multiphase case, we represent the community assignments $g$ in terms of an $N\times\nhat$ matrix whose $i,\al$ entry is $\delta_{g_i\al}$, where $\delta$ is the Kroneker delta. Instead of a double-well potential, we use a multi-well potential on $\R^{N\times\nhat}$ whose value is minimized by arguments with exactly one nonzero entry in each row. For example, Garcia-Cardona et al.~\cite{multi_mbo} proposed the following potential:
\begin{equation*}
	T(U) = \sum_{i=1}^N \left( \prod_{\alpha=1}^{\hat{n}} \frac{1}{4} \left\|U_i - e_k\right\|^2_{\ell^1} \right)\,,
\end{equation*}
where $U_i$ is the $i$th row of the $N\times\nhat$ matrix $U$ and $e_k$ is an $\hat{n}$-element vector that is equal to $0$ except for a $1$ in the $k$th entry.

For the particular case of surface-tension dynamics, we proceed as follows. Additionally, we assume in this subsection and the next that we have already eliminated the diagonal of $W$ (see~\cref{app:diag}). 

Given community assignments (and hence a partition of a network), if $U$ is the corresponding $N\times\hat{n}$ matrix, one can show that $W.*(U^TLU) = -W.*(U^TAU)$, where $.*$ is the entry-wise product.%
\footnote{As a proof, we note that $[W.*(U^T\diag(k)U)]_{\al\be} = \sum_i W_{\al\be}U_{i\al}k_i U_{i\be} = 0$, because $U_{i\al}U_{i\be}=0$ if $\al\ne\be$ and $W_{\al\al}=0$.}
Therefore, an appropriate GL functional for our problem is
\begin{equation}\label{gl_sbm}
  	\sumab \left[ -W_{\alpha\beta}U_\alpha^TLU_\beta + \frac{\vol(\al)\,\, e^{-W}_{\al\be}\,\,\vol(\be)}{2m}\right] + \sum_\al W_{\al\al}\vol(\al) + \frac{1}{2\epsilon}T(U)\,.
\end{equation}
Because $k^T U$ gives the vector of volumes, one can rewrite \cref{gl_sbm} as
\begin{equation}
  \sumab  \left[-W_{\alpha\beta}U_\alpha^TLU_\beta + \frac{k^T U_{\al} e^{-W}_{\al\be} U_\be^T k}{2m}\right] + \sum_\al W_{\al\al}\vol(\al) + \frac{1}{2\epsilon}T(U)\,,
	\label{exp:AC1}
\end{equation}
where $e^{-W}$ is the entry-wise exponential.

As in a continuum setting, one can prove $\Gamma$-convergence.
\begin{theorem}
  Let $W\in\R^{\nhat\times\nhat}$. The functionals in~\cref{exp:AC1} $\Gamma$-converge (as functions on $\R^{N\times\nhat}$) to~\cref{exp:cut_formulation} as $\epsilon\to 0$.
	\label{thm:gamma}
\end{theorem}
See~\cref{gam_proof} for a proof. As far as we are aware, this is the first $\Gamma$-convergence result for a multiphase graph energy on arbitrary graphs. However, see~\cite{braxton1,braxton2,braxton3} for $\Gamma$-convergence applied to consistency of multiphase geometric graph energies.

The resulting AC equation is
\begin{equation}\label{this_system}
  U_t = LUW - \frac{1}{2m}kk^TUe^{-W} - k .* \diag(W) - \frac{1}{\epsilon}T'(U)\,.
\end{equation}
 See~\cref{details} for further details on the numerical solution of \cref{this_system}.


\subsection{MBO Iteration}

\begin{algorithm}
	\begin{algorithmic}
		\State Input the initial domain.
		\State Initialize $u$ as the characteristic function of the initial domain.
		\For{$i=1,\ldots$}
		\State $u^{i+1/2}$ is the solution at time $\dt$ of $u_t = \Delta u$ with initial condition $u^i$\,.
		\State $u^{i+1} = \lfloor u^{i+1/2} + 0.5 \rfloor$\,, where $\lfloor \cdot\rfloor$ is the floor function. 
		\EndFor
		\State Output the set of points for which $u=1$\,.
	\end{algorithmic}
	\caption{A two-phase, continuum MBO scheme.
	}
	\label{alg:mbo}
\end{algorithm}

In~\cite{MBO_1992}, Merriman, Bence, and Osher showed that continuum MCF is well-approximated by the simple iteration in~\cref{alg:mbo}. In a rectangular domain, the iteration is extremely efficient, as one can use a fast Fourier transform when solving the heat equation. Esedoglu and Otto~\cite{esedoglu_otto} developed a generalized version of the MBO scheme (see ~\cref{alg:multi_mbo}) for computing the evolution of multiphase systems that are modeled by~\cref{exp:cont_ste}.

\begin{algorithm}
	\begin{algorithmic}
		\State Input the initial state of the domain.
		\State Initialize $u_1,\ldots,u_{\hat{n}}$ as the characteristic functions of the initial domains.
		\For{$i=1,\ldots$}
		\For{$\al=1,\ldots,\hat{n}$}
		\State $u^{i+1/2}_\alpha$ is the solution at time $\dt$ to $u_{\al,t} = \Delta u_\al$ with initial condition $u^i_\al$\,.
		\EndFor
		\For{each point $x$}
		\State $\hat{\al} = \argmin_{\al} \sum_{\be} \sigma_{\al\be}u_{\be}(x)\,.$ \quad // [Choose uniformly at random in case of a tie] 
		\State $u_{\hat{\al}}(x) = 1$ and $u_{\be}(x)=0$ if $\be \ne \hat{\al}$\,.
		\EndFor
		\EndFor
		\State Output $u$\,.
	\end{algorithmic}
	\caption{A multiphase, continuum MBO scheme.} 
	\label{alg:multi_mbo}
\end{algorithm}

One can apply the MBO idea to community detection in networks by replacing the continuum Laplacian with the (negative) combinatorial graph Laplacian, replacing $\sigma$ with $W$, changing $u$ to $U$, and adding appropriate forcing terms for the gradient descent of the volume-balance terms. 
See~\cref{details} for additional implementation details.


\subsection{Learning \texorpdfstring{$\w$}{omega}}

The MCF, AC, and MBO algorithms are able to produce a good partition of a network, given $W$, but they do not include a way to find $W$. A simple way to address this issue is to use an expectation-maximization (EM) algorithm, in which one alternates between solving for $g$ with fixed $W$ (using MCF, AC, or MBO) and solving for $W$ with fixed $g$. Given $g$, one can find a closed-form expression for the optimal $W$ by differentiating~\cref{exp:cut_formulation} with respect to any component of $W$ and setting the result to $0$~\cite{karrer_newman}. 

One must be careful, however, because the optimal $W_{\al\be}$ is infinite when $\Cut(\al,\be)=0$. This is problematic, because once one of the entries in $W$ is infinite, it prevents $g$ in subsequent iterations from taking any nonzero value of $\Cut(\al,\be)$; this gives bad results in our test examples. (See~\cref{empirical} for a discussion of these examples.) We address this issue by modifying the EM algorithm to reset all infinite values of $W$ to $1.1 \times W_{\text{max}}$, where $W_{\text{max}}$ is the largest non-infinite element of $W$ and $1.1$ is a (hand-tuned) parameter that allows moderate growth in $W$. 

We also need to address another practical issue for an EM approach to work. Specifically, the algorithm that we have described thus far in this section often finds bad local minima in which communities are merged erroneously or a single community is split inappropriately.\footnote{In other words, one can improve these local minima either by merging or by splitting existing communities. This is a special (and convenient, in this case) form of nonconvexity.}

To overcome this issue, we implement a wrapper function (see ~\cref{alg:wrapper}) that checks each community that is returned by MCF, AC, or MBO for further possible splitting or merging with other communities. Whenever we call MCF, AC, or MBO on a subgraph, we use the values of $k$ and $m$ for the whole graph rather than those for a subgraph. A similar idea was used in \cite{newman2006pre,hui} for their recursive partitioning procedures.

There is also a danger of overfitting by setting $\hat{n}=N$, which gives a likelihood of $1$ in~\cref{exp:SBM_MLE}. The proper selection of $\hat{n}$ is a complicated problem, both algorithmically and theoretically~\cite{reinert2016,reinert2017}. For our tests, we were very successful by using a simple heuristic approach. (Our framework is also compatible with more sophisticated methods for selecting $\hat{n}$.) For each data set, we supply an expected value of $\hat{n}$ for that data set, and we then add a quadratic penalty to the objective-function value whenever $\hat{n}$ differs from its expected value. This helps curtail overfitting, while still allowing our algorithms to perform merges and splits to escape bad local minima. Notably, this penalty does not alter the MCF, AC, or MBO procedures; instead, it is part of~\cref{alg:wrapper}.

\begin{algorithm}
	\begin{algorithmic}
		\State Input $A$, $\hat{n}_{\text{expected}}$\,.
		\State Place all nodes in the same community and add this community to a queue.
		\While{the queue is not empty}
		\State Save $g_{\text{old}}=g$ and $W_{\text{old}}=W$\,.
		\State Save the current objective-function value as $Q_{\text{old}}$\,.
		\State Partition the next community (as an induced subgraph, as we include all associated edges\footnotemark) in the queue 
		into $\min\{\hat{n}_{\text{expected}},\sqrt{N}\}$ communities using MCF, AC, or MBO with $w_{\al\be} = \begin{cases}
			1\,, & \al=\be\,, \\
			0.1\,, & \al\ne\be\,.
		      \end{cases}\,.$
		\While{it is possible to improve the objective-function value by merging two partition elements}
		\State Perform the merge that most improves the objective function.
		\EndWhile
		\If{the objective-function value is larger than $Q_{\text{old}}$}
		\State Add any newly created communities to the queue.
		\Else
		\State Set $g=g_{\text{old}}$ and $W=W_{\text{old}}$\,.
		\State Remove the current community from the queue.
		\EndIf
		\EndWhile
		\State Output $g$, $W$\,.
	\end{algorithmic}
	\caption{Our splitting--merging wrapper for escaping from bad local minima. In this algorithm, $Q$ is the objective function from~\cref{exp:cut_formulation} multiplied by $[1 + 0.1 \left( \nhat - \nhat_{\text{expected}} \right)^2]$, where we chose the value $0.1$ based on hand-tuning.
	}
	\label{alg:wrapper}
\end{algorithm}

\footnotetext{For this step, note that we use the degrees and number of edges from the entire graph, rather than the induced subgraph, when computing volumes. See~\cite{hui,boyd_modularity}, which make an analogous adjustment in the associated recursive step of their algorithms.}


\section{Empirical Results} \label{empirical}

We now discuss our results from several numerical experiments to (1) confirm that our algorithms can successfully recover $g$ and $\w$ from networks that we generate using SBMs and (2) explore their applicability to real-world networks.
In our experiments, we use three different families of SBMs, three Facebook networks (whose community structure is partly understood~\cite{fb,fblong}), and an example related to hyperspectral video segmentation. 
Because of the random initialization in our approach, we perform three trials on each of the networks for each algorithm, and we report the best result in each case.\footnote{We use three trials (using three different networks drawn from the random-graph models) to illustrate that our algorithms do not require a large number of attempts to reach a good optimum. In most of our trials, even a single run of a solver is likely to give good results. In~\cref{tab:main_table}, we report our best scores. Our worst scores for MCF are $0.00$, $0.00$, $0.00$, $-0.14$, and $0.01$ for the PP, MS, LFR, Caltech, and Princeton networks, respectively. (We did not record the worst score for Penn. St. or the plume network.) Our corresponding worst scores for AC and MBO are $0.00$, $0.00$, $0.01$, $0.22$, and $0.86$ (for AC) and $0.15$, $0.00$, $0.02$, $0.53$, and $1.12$ (for MBO). Comparing these results with~\cref{tab:main_table}, we see that our best and worst scores are often similar to each other.}
  For comparison, we also report the results of a Kernighan--Lin (KL) algorithm, which was reported in~\cite{karrer_newman} to be effective. We summarize our results in~\cref{tab:main_table}, and we highlight that we consistently recover the underlying structure in the synthetic examples. For the real networks, we compare our results with a reference partition based on metadata that is thought to be correlated with the community structure.
We find that the MCF scheme performs the best among our three schemes on these networks, and it finds partitions with a larger likelihood than the reference partition.\footnote{In synthetic networks, the reference partitions represent a ``ground truth,'' in the sense that they reflect the principle upon which we constructed the network. For these networks, finding a partition with higher likelihood than the reference partition reflects the fact that the data is stochastic, and a maximum-likelihood partition may differ slightly from the ground-truth one. An algorithm that achieves a likelihood that is higher than ground truth is more successful at optimizing the likelihood function than one that does not. In real networks, the reference partition is not a ``ground truth.'' Instead, it is a point of reference that is based on a ``natural'' grouping of the nodes when one is available. In most applications of community detection, there is no ground truth \cite{peel}. Real networks can have many different organizing principles and multiple insightful partitions, including both (1) partitions that are slight variations of each other that yield similar values of objective functions and (2) partitions that are very different from each other that yield similar values of such functions \cite{good_2010}. In particular, the fact that some of our methods find partitions with higher likelihood scores than the reference partition is not indicative of a failure of the maximum-likelihood approach, because there is no reason for a reference partition to be the best possible partition of a network. (For the Facebook networks, for example, it is known that this is not the case \cite{hric2016metadata}.) To the extent that SBM MLE is appropriate for the data and application, our partitions are sometimes better than the reference partitions.}
We implement our methods in {\sc Matlab}, so one should interpret computation times in~\cref{tab:timing} as indicative that the run time is reasonable for networks with millions of edges. Given a careful implementation in a compiled language, it is possible to study even larger networks.\footnote{One may perhaps construe from Table \ref{tab:timing} that the MCF method's speed on the large plume example indicates that its scaling is better than linear with the number of nodes, which is of course not true, because we alter each node during each iteration in our implementation. Several factors influence the observed computation times. For example, larger data improves the effectiveness of parallelization and vectorization of operations (which {\sc Matlab} does automatically for certain operations). Furthermore, the operation count can be quadratic in the number of communities, which is heterogeneous across the data sets that we examined and is largest in the LFR example. Finally, the different networks are very different structurally, which affects the number of iterations that are necessary for convergence.}
For an example of code for a similar problem that was solved by an MBO scheme at large scale (including a weighted graph with almost $14$ million nodes and $1.8\times 10^{14}$ edges), see~\cite{gloria}.

We briefly describe the three families of SBM-related networks that we use in our numerical experiments.
\begin{itemize}
	\item Planted partition (PP) is a 16,000-node graph that consists of $10$ equal-size communities. It is produced by the method that was described in~\cite{karrer_newman}. It builds a degree-corrected SBM with a truncated power-law degree distribution with exponent $2$. The parameter $\lambda$ from Equation~(27) in~\cite{karrer_newman} is $0.001$, indicating a fairly clear separation between communities.
	\item Lancichinetti--Fortunato--Radicchi (LFR) is a standard benchmark SBM network \cite{LFR_2008}. We construct 1000-node LFR graphs with a power-law degree distribution (with exponent $2$), mean degree $20$, maximum degree $50$, power-law-distributed community sizes (with exponent $1$), community sizes between $10$ and $50$ nodes, and mixing parameter $0.1$. 
	\item Multiscale SBM (MS). To construct such a graph, we take a sequence of disjoint components; in order, these are a $10$-clique, a $20$-clique, and a sequence of Erd\H{o}s--R\'enyi (ER) graphs (drawn from the $G(n,p)$ model with $n$ nodes and $np = 20$) of sizes $40$, $80$, $160$, \ldots, $5120$. Each of these graphs has a total of $10,230$ nodes. In each such graph, we connect the components to each other by adding a single edge, from nodes chosen uniformly at random, between each consecutive clique or ER graph. This construction tests whether an algorithm can find communities of widely varying sizes in the same graph~\cite{fortunato_barthelemy_2007, arenas}.
\end{itemize}

\begin{figure}
	\centering
	\includegraphics[width=.5\textwidth]{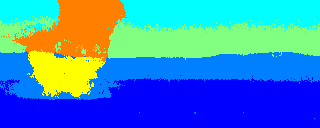}
	\caption{Segmentation of a hyperspectral video using graph MCF. The gas plume is clearly represented in the yellow and orange pixels. The two blue communities on the bottom are the ground, and the other two communities are the sky. This image is frame 3 of 7. (It is best to view this plot in color.)}
	\label{fig:plume}
\end{figure}

The hyperspectral video is a recording of a gas plume as it was released at the Dugway Proving Ground~\cite{gerhart2013detection,plumes_source,merkurjev_hyperspectral_2014}. A hyperspectral video is different from an RGB video, in that each pixel in the former encodes the intensity of light at a large number (e.g., $129$, in this case) of different wavelengths rather than at only $3$, with each channel corresponding to a wavelength. We consider the classification problem of identifying pixels that include similar materials (such as dirt, road, grass, and so on). This problem is difficult because of the diffuse nature of the gas, which leads to a faint signal that spreads out among many wavelengths and with boundaries that are difficult to determine. We construct a graph representation of this video using ``nonlocal means,'' as described in~\cite{buades_2005}. Specifically, we use the following construction. For each pixel $p$ and in each of $7$ frames, we construct a vector $v_p$ by concatenating the data in a $3\times3$ window that is centered at $p$. We then use a weighted cosine similarity measure (which is a common choice for hyperspectral imaging applications) on these $(3\times3\times 129)$-component vectors, where we give the most weight to the components from the center of the window.\footnote{We weight the center pixel components by $1$, the components from adjacent pixels by $0.5$, and the components from corner pixels by $0.25$. That is, we let $v_{ij}$ be the $129$-element vector at pixel $(i,j)$, and we define $w_{ij}$ as the concatenation of $v_{ij}$, $.5 v_{i+1,j}$, $.5 v_{i-1,j}$, $.5 v_{i,j+1}$, $.5 v_{i,j-1}$, $.25 v_{i+1,j+1}$, $.25 v_{i+1,j-1}$, $.25 v_{i-1,j+1}$, and $.25 v_{i-1,j-1}$. We then calculate the cosine similarity between each pair of $w_{ij}$ vectors.} Finally, using the {\sc VLFeat} software package~\cite{vlfeat}, we build an unweighted 10-nearest-neighbor graph using the similarity measure and a $k$-dimensional tree (with $k=10$)~\cite{kdtree}. We see from~\cref{fig:plume} that partitions with small values of~\cref{exp:cut_formulation} correspond to meaningful segmentations of the image.

\begin{table}
	\setlength\tabcolsep{5.5pt} 
	\begin{tabular}{l|llllllll}
		\toprule
		&	&	{PP}	&	{LFR}	&	{MS}	&   {Caltech}	&	{Princeton}	&  {Penn. St.}	&    {Plume}	\\
		\midrule                                                                                 
		&Nodes	&16,000		&	1,000	&	10,230	&	762	&	6,575		&	41,536	&      284,481	\\
		&Edges	&$2.9\times 10^5$&$9.8\times 10^3$&$1.0\times 10^5$&$16,651$	&$293,307$	& $1,362,220$	&	$2,723,840$\\
		&Communities& 	10	&	40	&	10	&	8	&	4	&	8	&	5	\\
		\midrule
		\multirow{5}{*}{Score}
		&MCF 	&0	&	0	&	0 	&	$-0.16$ &	 $-0.02$	&  	$-0.56$	&	$-1.41$	\\
		&AC 	&0	&	0	&	0 	&	0.21 	&	 0.58		&	$-0.04$	&	$-1.23$	\\
		&MBO 	&0	&	0	&  	0 	&	0.53	&	 1.12		&	0.40	&	$-1.21$	\\
		&KL	&0.28	&	0.03	&  	0.04	&	$-0.16$ &   	 0.11		&	$-0.55$	&	$-1.38$	\\
		&Reference&0	&	0	&	0 	&	0 	&	 0		&	0 	&	0	\\
		\bottomrule 
	\end{tabular}
	\caption{
		Results of several tests on several synthetic and empirical networks. We use three surface-tension-based methods (mean-curvature flow, Allen--Cahn, and Merriman--Bence--Osher) and the Kernighan--Lin algorithm from~\cite{karrer_newman} to partition three synthetic networks (Planted Partition, LFR, and Multiscale SBM) and the largest connected components of three empirical networks (Caltech36, Princeton12, and Penn94) from the {\sc Facebook100} data set \cite{fblong}. The score is the recovered surface-tension energy~\cref{exp:cut_formulation} minus the corresponding energy of a reference partition, divided by the absolute value of the energy of the reference partition. Smaller values indicate better performance, and $0$ corresponds to a partition that is of comparable quality as the reference partition. For the synthetic networks, we use the planted (and hence ground-truth) community structure as the reference partition. For the Facebook networks, we use metadata that is positively correlated with community structure (namely, House affiliation for Caltech and graduation year for the other two networks). For the plume video, our reference partition is to assign all nodes to the same community, because no pixel-level metadata is associated with the images. The edge counts on the synthetic networks constitute an order-of-magnitude approximation, because the exact number differs across the three instantiations of these models. (This table gives the best result among three tests for each example; for the random-graph models, each such example is a different graph from the same model.)
	} 
	\label{tab:main_table}
	\setlength\tabcolsep{6pt} 
\end{table}

\begin{table}
	\centering
	\begin{tabular}{|r|r|r|r|r|r|r|r|}
		\hline
		& PP 	& LFR	& MS 	& Caltech & Princeton 	& Penn. State 	&	Plume \\ 
		\hline
		\hline
		MCF 	& 5.36 	& 17.71	& 3.47	& 1.39	  & 1.46	& 38.91		&	77.91\\
		AC 	& 5.37	& 26.27	& 7.28	& 8.84	  & 480.4	& 3853 		&	268.7 \\
		MBO 	& 4.27	& 11.05	& 1.73	& 0.67	  & 7.43	& 382.31	&	270.0  \\
		KL 	& 16,566& 176	& 5,117	& 20	  & 662		& 95,603  	&	980,520 \\
		\hline
	\end{tabular}
	\caption{Median computation times (in seconds) for our example networks.
	}
	\label{tab:timing}
\end{table}

In~\cref{tab:sigma}, we show an example of a $W$ matrix that we obtain from an MS network to illustrate that we recover different surface tensions between different pairs of communities.\footnote{For this example, we used the change of variables from~\cref{app:diag} to eliminate the diagonal elements.}

\begin{table}
	\centering
	\begin{tabular}{llllllllll}
		0&5.22&$\infty$&$\infty$&$\infty$&$\infty$&$\infty$&$\infty$&$\infty$&$\infty$\\
		5.22&0&6.1817&$\infty$&$\infty$&$\infty$&$\infty$&$\infty$&$\infty$&$\infty$\\
		$\infty$&6.1817&0&6.8471&$\infty$&$\infty$&$\infty$&$\infty$&$\infty$&$\infty$\\
		$\infty$&$\infty$&6.8471&0&7.6316&$\infty$&$\infty$&$\infty$&$\infty$&$\infty$\\
		$\infty$&$\infty$&$\infty$&7.6316&0&8.362&$\infty$&$\infty$&$\infty$&$\infty$\\
		$\infty$&$\infty$&$\infty$&$\infty$&8.362&0&9.0869&$\infty$&$\infty$&$\infty$\\
		$\infty$&$\infty$&$\infty$&$\infty$&$\infty$&9.0869&0&9.7926&$\infty$&$\infty$\\
		$\infty$&$\infty$&$\infty$&$\infty$&$\infty$&$\infty$&9.7926&0&10.4911&$\infty$\\
		$\infty$&$\infty$&$\infty$&$\infty$&$\infty$&$\infty$&$\infty$&10.4911&0&11.1869\\
		$\infty$&$\infty$&$\infty$&$\infty$&$\infty$&$\infty$&$\infty$&$\infty$&11.1869&0\\
	\end{tabular}
	\caption{Optimal surface tensions for the MS SBM example. (The entries are heterogeneous, so there are different surface tensions between different pairs of communities.) The infinite entries correspond to sets with no observed edge between them.}
	\label{tab:sigma}
\end{table}


\section{Conclusions and Discussion}\label{conc}

We have shown that a particular stochastic block model (SBM) maximum-likelihood estimation (MLE) problem is equivalent to a discrete version of a well-known surface-tension problem. This equivalence, which associates graph cuts to surface areas and SBM parameters to physical surface tensions, gives new geometric and physical interpretations to SBM MLE problems, which are traditionally viewed from a statistical perspective. We used the new connection to adapt three well-known surface-tension-minimization algorithms to community detection in graphs. Our subsequent computations suggest that the resulting algorithms are able to successfully find underlying community structure in SBM-related graphs. When applied to graphs that are constructed from empirical data, our mean-curvature-flow (MCF) method performs very well, but the other two methods face some issues (which will be interesting to explore in future studies).\footnote{Several factors seem to contribute to the performance difference, and it is impossible to disentangle them without extensive additional testing. The following are some possible contributing factors. The Facebook networks are not generated from an SBM, and the former have more complicated structures than those in synthetic networks. Our particular solution method involving eigenvectors may thus be less appropriate for solving diffusion equations in the Facebook networks than in synthetic networks. It is empirically clear that the Facebook networks have more eigenvector localization (e.g., as measured using inverse participation ratio) than the synthetic ones. The AC and MBO methods deal with the nonconvexity differently than MCF. (The former two use pseudospectral methods to jump to better regions, whereas MCF allows all boundary nodes to move independently and simultaneously.)
It would be interesting to conduct a detailed study of these various factors using a large variety of networks, as it will likely improve scientific understanding of the geometry and associated flows for different families of networks.}
We also proved a $\Gamma$-convergence result that gives theoretical justification for our algorithms.

Although our paper has focused on a specific form of an SBM and an associated MLE problem, our techniques should also be insightful for other studies of SBMs and their applications. One straightforward adaption is to consider SBMs without degree correction, although that is more interesting for theoretical work than for applications. Additionally, it seems promising to incorporate priors on the values of $g$ and $\w$ as regularizers in the surface-tension energy (perhaps in a way that is similar to the procedure in~\cite{luo_comparison}). Another viable extension is to incorporate a small amount of supervision into the community-inference process using techniques (such as quadratic fidelity terms) from image processing. A similar idea was used for modularity maximization in~\cite{hui} and was tested further in~\cite{boyd_modularity}. 

Introducing supervision helps alleviate severe nonconvexity by penalizing local minima that are inconsistent with the (ideally) ground-truth classifications from which one draws the supervision. It is also important to generalize our approach to more complicated types of networks, such as multilayer \cite{kivela2014} and temporal networks \cite{holme2012}, and to incorporate metadata \cite{newman-clauset2016} into our inference methodology.
For example, given our successful results on the hyperspectral video, it may be particularly interesting to use temporal network clustering to analyze time-dependent communities in the video.

Approaches such as inference using SBMs and modularity maximization are also related to other approaches for community detection, and the results in the present paper may help further illuminate those connections. These include recent work that relates SBMs to local methods for community detection that are based on personalized PageRank \cite{Kloumann33} and very recent work that established new connections between modularity maximization and several other approaches \cite{gleich2017}. We expect that further mapping of the relations between the diverse available perspectives for community detection (and other problems in network clustering) will yield many new insights for network theory, algorithms, and applications.


\section*{Acknowledgements}

ZMB and ALB were funded by NSF grants DMS-1737770 and DMS-1417674, as well as ONR grant N00014-16-1-2119. ZMB was also supported by the Department of Defense (DoD) through the National Defense Science \& Engineering Graduate Fellowship (NDSEG) Program and the Eunice Kennedy Shriver National Institute of Child Health \& Human Development of the National Institutes of Health under Award Number R01HD075712. All three authors were supported by DARPA award number FA8750-18-2-0066. We thank Brent Edmunds, Robert Hannah, and  Kevin Miller for helpful discussions. We also thank two anonymous referees for many helpful comments.

The content is solely the responsibility of the authors and does not necessarily represent the official views of any of the agencies that supported this work.


\begin{appendices}
  \crefalias{section}{app} 


  \section{Eliminating the Diagonal Elements of \texorpdfstring{$W$}{W}} \label{app:diag}

It is difficult to interpret the parameters $W_{\al\al}$ in the context of~\cref{exp:cut_formulation} and our surface-tension analogy, because they correspond to ``internal'' surface tensions of a single crystal. In this appendix, we use a change of variables to eliminate these diagonal terms and replace them with additional volume terms, which are much easier to interpret.

We begin with the identity
	\begin{align}
		\sumab  W_{\alpha\beta} \Cut(\alpha,\beta) = \sum_\alpha \sum_{\beta \ne \alpha} W_{\alpha\beta} \Cut(\alpha,\beta) + \sum_{\alpha} W_{\alpha\alpha} \Cut(\alpha,\alpha)\,,
		\label{eqn:with_diag}
	\end{align}
	and we compute
	\begin{align}
		\sum_\alpha W_{\alpha\alpha} \Cut(\alpha,\alpha) &= \sum_\alpha W_{\alpha,\alpha} \sum_{ g_i=\alpha,g_j=\alpha} w_{ij} \notag  \\
		&= \sum_\alpha W_{\alpha,\alpha} \left(\sum_{ g_i=\alpha,j=1,\dots,N} w_{ij} - \sum_{g_i=\alpha,g_j\ne\alpha} w_{ij} \right) \notag \\
		&= \sum_\alpha W_{\alpha,\alpha} \left(\sum_{ g_i=\alpha} k_i - \sum_{\beta\neq\alpha,g_i=\alpha,g_j=\beta} w_{ij} \right) \notag \\
		&= \sum_\alpha W_{\alpha,\alpha} \left(\vol(\alpha)- \sum_{\beta\neq\alpha} \Cut(\alpha,\beta) \right)\,.
		\label{eqn:diag}
	\end{align}
Combining~\cref{eqn:diag} with~\cref{eqn:with_diag} yields
	\begin{align}
	  \sumab  W_{\alpha\beta} \Cut(\alpha,\beta) = \sum_{\alpha\ne\beta} \left(W_{\alpha\beta}-W_{\alpha\alpha}\right) \Cut(\alpha,\beta) + \sum_\alpha W_{\alpha\alpha} \vol(\alpha)\,,
		\label{eqn:without_diag}
	\end{align}
	assuming\footnote{The case in which $W_{\al\al}=\infty$ does not occur in our methods.} that $W_{\al\al}$ is finite for each $\al$.
	This formulation removes the diagonal from the double sum at the cost of introducing asymmetry into the subscripts of the coefficients.
	We can fix this new issue by replacing~\cref{eqn:without_diag} with
\begin{align}
		\sumab  W_{\alpha\beta} \Cut(\alpha,\beta) 
		&= \sum_{\alpha\ne\beta} \left(W_{\alpha\beta}-\frac{1}{2} W_{\alpha\alpha} - \frac{1}{2} W_{\beta\beta} \right) \Cut(\alpha,\beta) + \sum_\alpha W_{\alpha\alpha} \vol(\alpha) \notag \\
		&= \sum_{\alpha\ne\beta} \hat{\sigma}_{\alpha\beta} \Cut(\alpha,\beta) + \sum_\alpha W_{\alpha\alpha} \vol(\alpha)\,,
		\label{eqn:symmetrized}
\end{align}
where $\hat{\sigma}_{\al\be} = W_{\al\be} - \frac{1}{2}W_{\al\al} - \frac{1}{2}W_{\be\be}$. The matrix $\hat{\sigma}$ is symmetric and has $0$ values on the diagonal.

Finally, we expand a bit on the role of the volume terms in~\cref{exp:cut_formulation}. The term
\begin{equation}\label{latter}
	\sum_\alpha W_{\alpha\alpha} \vol(\alpha)
\end{equation}
is the inner product of the vector of volumes with the diagonal of $W$. We minimize~\cref{latter}, subject to the constraints $\sum_{\alpha} \vol(\al) = 2m$ and $\vol(\al)\ge 0$, by placing all of the nodes in the community that corresponds to the smallest\footnote{When referring to ``smallest'' eigenvalues in the appendices, we mean the smallest positive or most-negative values rather than those that are smallest in magnitude.} entry in the diagonal of $W$. Therefore, these terms incentivize placing more mass in the communities that have the smallest volume penalties.


\section{\texorpdfstring{$\Gamma$}{Gamma}-Convergence of the Ginzburg--Landau Approximation of \texorpdfstring{\eqref{exp:cut_formulation}}{Graph Surface Tension Energy}}
\label{gam_proof}

The notion of $\Gamma$-convergence is defined as follows:
	\begin{definition}
		Let $Y$ be a metric space, and let $F_n$ be a sequence of functionals that take values in $\R\cup \{\infty\} \cup \{-\infty\}.$ We say that $F_n$ \emph{$\Gamma$-converges} to another functional $F$ if for all $x\in Y$, the following bounds hold:
		\begin{enumerate}
			\item (Lower bound) For every sequence $x_n\to x$, we have $F(x)\le \liminf_{n\to\infty} F_n(x_n)$\,.
			\item (Upper bound) For every $x\in Y$, there is a sequence $x_n\to x$ such that $F(x)\ge \limsup_{n\to\infty} F_n(x_n)$\,.
		\end{enumerate}
	\end{definition}
We now prove~\cref{thm:gamma}.
\begin{proof}
We largely follow~\cite{van_gennip_2012}, although we generalize to account for the multiphase nature of our problem.

The terms that do not involve the potential $T$ are continuous and independent of $\epsilon$, so they cannot interfere with $\Gamma$-convergence~\cite{dal_maso}.\footnote{The graph-TV term is a composition of addition, subtraction, projection onto components, and taking absolute values; therefore, it is continuous.} Consequently, it suffices to prove that $\frac{1}{\epsilon}T : \mathbb{R}^{N\times\nhat}\to\R$ $\Gamma$-converges to 
\begin{equation*}
			\chi(U) = 
			\begin{cases}
				0\,, & \text{if $U$ corresponds to a partition}\,, \\
				+\infty\,, & \text{otherwise}\,.
			\end{cases}
\end{equation*}

To prove the lower bound, let $U_n \to U$ and $\epsilon_n\to 0$. (In this proof, the subscript $n$ indexes the sequence, rather than the matrix columns.) If $U$ corresponds to a partition, $\chi(U)=0$, which is automatically less than or equal to $\frac{1}{\epsilon_n}T(U_n)$ for each $n$. If $U$ does not correspond to a partition, $\chi(U) = +\infty.$ There exists a constant $c>0$ such that the distance (in, for example, the Frobenius norm) from $U_n$ to the nearest feasible point (i.e., a point corresponding to a partition) is at least $c$ as $n\to\infty$. Let $T_c$ be the infimum of $T$ on all of $\R^{N\times\hat{n}}$ except for the balls of radius $c$ that surround each feasible point (so, in particular, $T_0>0$). It follows that $\lim\inf_{n\to\infty} \frac{1}{\epsilon_n}T(U_n) \ge \lim_{n\to\infty} \frac{1}{\epsilon_n}T_0 = +\infty$. Therefore, the lower bound always holds.

To prove the upper bound, let $U$ be any $N\times\hat{n}$ matrix. If $U$ corresponds to a partition, then letting $U_n=U$ for all $n$ gives the required sequence. If $u$ does not correspond to a partition, then $U_n=U$ for all $n$ still satisfies the upper bound.

Therefore, both the upper and lower bound requirements hold, and we have proven $\Gamma$-convergence.
\end{proof}


\section{Additional Notes on the AC and MBO Schemes} \label{details}

In this appendix, we discuss some practical details about our implementation of the AC and MBO solvers.

The choice of $\epsilon$ in AC is important, because it selects a characteristic scale of the transition between the $U_\al \approx 1$ and $U_\al \approx 0$ regions. If $\epsilon$ is too small, the barrier to transition is large, and no evolution occurs. If it is too large, the transition layer includes so many nodes that $U$ does not approximately correspond to a partition of a graph. Furthermore,~\cref{thm:gamma} asserts only that the minimizers of~\cref{exp:cut_formulation} and~\cref{exp:AC1} are related when $\epsilon$ is sufficiently small. In our numerical experiments, we set $\epsilon = 0.004$, a choice that we selected by hand-tuning using our synthetic networks. There is no reason to believe that the same value should work for all networks. For example, for the well-known Zachary Karate Club network \cite{karate}, we obtain much better results for $\epsilon=0.04$. A very interesting problem is to determine a correct notion of distance and accompanying quantitative estimates to allow an automated selection of $\epsilon$ to obtain a transition layer with an appropriate width to give useful results. We discretize the AC equation via convex splitting~\cite{eyre}:
\begin{equation*}
  (1 + c \,\dt)U^{n+1} - LU^{n+1}W = -\dt\,\left(cU^n + k .*\diag(W)+T'(U^n) + \frac{1}{2m}kk^TUe^{-W}\right)\,,
\end{equation*}
where $c > {2}/{\epsilon}$~\cite{luo}. Using the constant $c$ leads to an unconditionally stable scheme, which negates the stiffness caused by the ${1}/{\epsilon}$ scale.

It is necessary to solve a linear system of the form
\begin{equation}\label{linear}
	(1 + c \,\dt)U^{n+1} - LU^{n+1}W = F^n
\end{equation}
many times. In a continuum setting, one can use a fast Fourier transform, but we do not know of a graph analog with comparable computational efficiency. Instead, we find the $2\hat{n}$ eigenvectors that correspond to the smallest eigenvalues\footnote{The number $2\hat{n}$ is somewhat arbitrary; we choose it to exceed $\hat{n}$, but for computational convenience, we do not want it to be too large.} of $L$ and the entire spectrum of $W$. Therefore, $L$ is approximated by $V_LD_LV_L^T$, where $D_L$ is a $2\nhat\times 2\nhat$ diagonal matrix of the smallest eigenvectors of $L$, sorted from smallest to largest, and $V_L$ is the associated matrix of eigenvectors. Furthermore, let $W = V_WD_WV_W^T$ be the full spectral decomposition of $W$. The system~\cref{linear} is then approximately equivalent to
\begin{equation*}
	(1 + c \,\dt)V_L^TU^{n+1}V_W - D_LV_L^TU^{n+1}V_WD_W = V_L^TF^nV_W\,.
\end{equation*}
Letting $\hat{U}^n = V_L^TU^{n}V_W$ and $\hat{F}^n=V_L^TF^nV_W$, we write
\begin{equation} \label{linear3}
	(1 + c\, \dt)\hat{U}^{n+1} - D_L\hat{U}^{n+1}D_W = \hat{F}^n\,,
\end{equation}
which is easy to solve for $\hat{U}^{n+1}$. We convert $\hat{U}^{n+1}$ to a solution using $U^{n+1} = V_L\hat{U}^{N+1}V_W^T$. (See~\cite{bertozzi_flenner_2012} for a discussion of this method of recovering $U^{n+1}$ from $\hat{U}^{n+1}$.)

One final detail that we wish to note is that we want the evolution of $U$ to be restricted to have a row sum of $1$, so that we can interpret it in terms of probabilities. To do this, we use a modification of the projection algorithm from~\cite{chen_ye} at each time step.\footnote{The algorithm from~\cite{chen_ye} acts on a single row vector, and our modification is simply to process all rows at once by replacing operations on row-vector components with operations on matrix columns. The result is mathematically equivalent (up to round-off errors), but it is much faster because it vectorizes the operations.}

The MBO solver uses a very similar pseudospectral scheme, although it does not include convex splitting. Unlike in the AC scheme, we need to estimate two time steps automatically in our code, instead of tuning them by hand.\footnote{Tuning by hand is not only laborious, but it also is very problematic in cases involving recursion or when $W$ changes, because the correct time step depends both on the (sub)graph being partitioned and on $W$. Consequently, there may be no time step that works for all subgraphs and choices of $W$ even for a single data set.} The first is the inner-loop step (i.e., the time step that we use for computing the diffusion), which we determine using a restriction (which one can show is necessary for stability\footnote{See, e.g., Section 8.6 of ~\cite{numerical_ode} for a description of the necessary techniques, which are standard in the numerical analysis of ordinary differential equations.}) that the time step should not exceed twice the reciprocal of the largest eigenvalue of the linear operator that maps $U \to \frac{1}{m}kk^TUe^{-W}$. The time step between thresholdings of $U$ is given by the reciprocal of the geometric mean of the largest and smallest eigenvalues of the operator that maps $U\to LUW$. The associated intuition is that linear diffusion should have enough time to evolve (to avoid getting stuck) but not enough time to evolve to steady state (because the steady state does not depend on the initial condition, so it carries no information about it). The reciprocal of the smallest eigenvalue gives an estimate of the time that it takes to reach steady state, and the reciprocal of the largest eigenvalue gives an estimate of the fastest evolution of the system. We choose the geometric mean between these values to produce a number between these two extremes.\footnote{From experimentation, we concluded that it is better to multiply this time step by $8$ to avoid getting stuck too early. We chose the geometric mean, because eigenvalues can have very different orders of magnitude.} 
	References~\cite{boyd_modularity} and~\cite{van_gennip_2014} proved bounds (although in a simpler setting) that support these time-step choices for MBO schemes.





\end{appendices}

\end{document}